\documentclass[11pt]{article}
\newcommand{\topic}[1]{#1}
\usepackage[preprint]{acl}

\usepackage{booktabs}   %
\usepackage{tabularx}   %
\usepackage{caption}    %
\usepackage{times}
\usepackage{latexsym}
\usepackage{amsmath}
\usepackage{listings}
\usepackage{xcolor}
\usepackage{booktabs}
\usepackage{tabularx}
\usepackage{balance}
\usepackage{enumitem}
\usepackage{booktabs}
\usepackage{ragged2e}   
\usepackage{tabularx}

\definecolor{codegray}{rgb}{0.5,0.5,0.5}
\definecolor{backcolour}{rgb}{0.95,0.95,0.92}

\lstdefinelanguage{json}{
    basicstyle=\small\ttfamily,
    numbers=left,
    numberstyle=\tiny\color{codegray},
    stepnumber=1,
    numbersep=8pt,
    showstringspaces=false,
    breaklines=true,
    frame=lines,
    backgroundcolor=\color{backcolour},
    stringstyle=\color{blue},
    keywordstyle=\color{magenta}
}
\usepackage[T1]{fontenc}

\usepackage[utf8]{inputenc}

\usepackage{microtype}

\usepackage{inconsolata}

\usepackage{graphicx}
\usepackage{amssymb}

\title{Aura: Dynamic Intra-Turn Emotion-Aware Adaptation of \\ Large Language Model Responses}

\author{Rachel Schuchert \\
  Department of Computer Science \\
  ETH Zürich, Switzerland \\
  \And
  Christian Holz \\
  Department of Computer Science \\
  ETH Zürich, Switzerland \\
  }

\usepackage{booktabs}

\begin{document}
\maketitle
\begin{abstract}

Effective human-AI interaction requires systems that dynamically adapt to a user's behavior and evolving understanding.
When users interact with Large Language Models (LLMs), these models typically respond to prompts without sensing the user's \emph{immediate} reactions.
This lack of communicative synchrony can lead to information overload or leave confusion unresolved in real time.
In this paper, we introduce \emph{Aura}, a framework that enables LLM systems to dynamically modulate output based on a user's evolving emotions.
Aura's Perception Module continuously estimates the user's emotional state from facial expressions.
Our Policy Module then selects interventions through a probabilistic belief model.
Finally, Aura's Generation Module uses parameter-efficient Low-Rank Adaptation (LoRA) adapters to produce contextually tailored responses \emph{mid-turn} during response generation.
We evaluated Aura in a within-subjects user study ($N=20$) on information-seeking tasks, where it achieved statistically significantly higher normalized perceived learning gains than a Llama-3 baseline and reduced interaction time by 21\% relative to existing LLM baselines (GPT-4o, Llama-3).
Our results indicate that real-time, context-sensitive interventions can improve learning efficiency and user satisfaction without observable degradation in factual accuracy.
Aura thus supports the potential for more responsive and effective human-AI interaction.

\end{abstract}

\section{Introduction}
\begin{figure*}[t]
  \centering
\includegraphics[
    width=\textwidth,
     trim=6.5cm 6.3cm 6.5cm 5.9cm, %
     clip
  ]{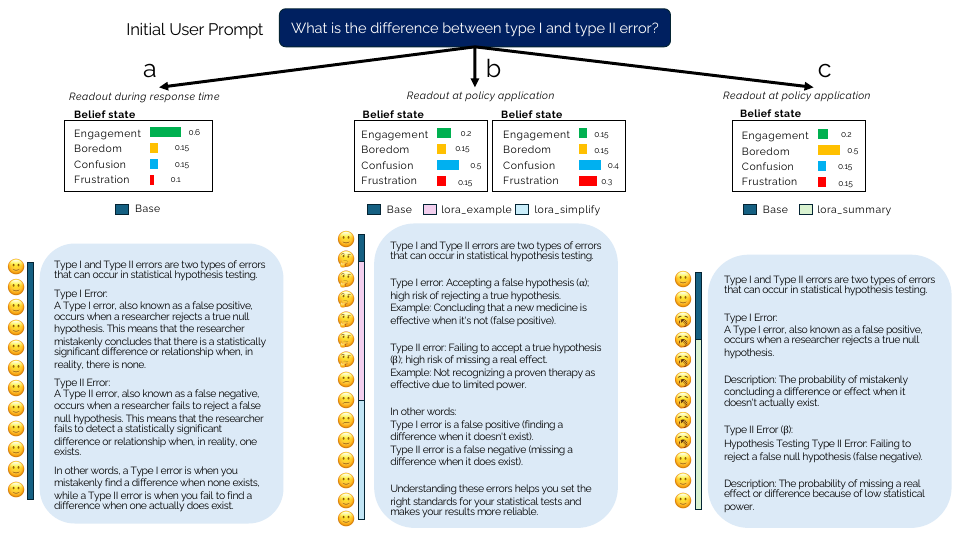}
  \caption{Three distinct interactions: 
  (a) \textbf{Sustained Engagement}: The Perception Module detects high engagement, leading the framework to continue with the base model for standard delivery; 
  (b) \textbf{Interaction Recovery}: Increasing inferred confusion and frustration within the belief state trigger mid-turn shifts to the Clarify (\texttt{lora\_example}) and Simplify (\texttt{lora\_simplify}) adapters to resolve knowledge gaps; 
  (c) \textbf{Efficiency Optimization}: Detected boredom activates the Pace (\texttt{lora\_summary}) adapter to condense information and maintain interaction velocity. }
  \label{fig:scenarios}
\end{figure*}

Human communication is guided by rapid emotional and behavioral signals that reflect evolving expectations and satisfaction~\cite{clark1991grounding, ekman1992argument}.
In interpersonal dialogue, speakers rely on this continuous stream of facial expressions, gaze, and posture to monitor a listener’s state of understanding~\cite{picard1997affective}.
These cues allow a speaker to naturally adjust their explanations in real time. %

In contrast, interaction with large language models (LLMs) is characterized through rather static prompt--response exchanges.
A user provides a textual prompt expressing an information need, and the model generates a complete response in a single, open-loop pass. 
A limitation of this interaction is that the model remains ``blind'' to the user \emph{during} the generation process, which results in requiring more interaction turns.
Users can communicate their reaction and a potential need for clarification due to confusion, frustration, or boredom in a subsequent interaction, if they express it at all.
Prior work has extended this by incorporating user context, such as long-term preferences or explicit personas, yet generation remains largely insensitive to \emph{immediate} feedback~\cite{zhong2023memory, deshpande2023toxicity}.
Existing adaptation strategies typically rely on explicit user instructions, pre-defined preferences, or iterative post-hoc edits~\cite{fan2018hierarchical, ouyang2022training, madaan2023self}.
While these methods allow LLMs to adapt their output \emph{between} individual turns of interaction, they fail to achieve the fluid, intra-turn synchrony common \emph{during} human-to-human communication.

In this paper, we introduce \emph{Aura}, a method for emotion-aware response adaptation that bridges the gap between static generation and dynamic human feedback.
Aura immediately adapts LLM output  to a person's spontaneously expressed facial expressions \emph{during} response generation rather than only between conversational turns.
Our system estimates a person's emotions via observing their facial expressions with a webcam to assess the person's implicit and immediate feedback on the suitability of the model's output, 
integrated temporally to maintain an evolving probabilistic estimate of user emotional state from initially noisy facial geometry cues.
Aura thus produces dynamically adaptive interventions that remain stable under uncertainty while preserving conversational continuity.
As a result, Aura can maintain engagement, address moments of confusion, and improve interaction efficiency in real time as shown in Figure~\ref{fig:scenarios}.

Aura integrates a Partially Observable Markov Decision Process (POMDP) model to track state transitions over time.
Based on estimated states, Aura selects specific Low-Rank Adaptation (LoRA) adapters that modulate detail, examples, and response termination for a frozen base LLM that allow low-latency, parameter-space control that is not achievable through prompting alone~\cite{kaelbling1998pomdp, young2013pomdp}.

By integrating user state modeling directly into the model's parameter space, Aura enables fine-grained, intra-turn adaptation, such as switching to simpler language or providing examples mid-response.
This work unifies real-time user state computing with inference-time control, providing a step toward more fluid, synchronized dialogue.

\begin{figure*}[t]
  \centering
\includegraphics[
    width=\textwidth,
     trim=.5cm 6cm 2.5cm 5cm, %
     clip
  ]{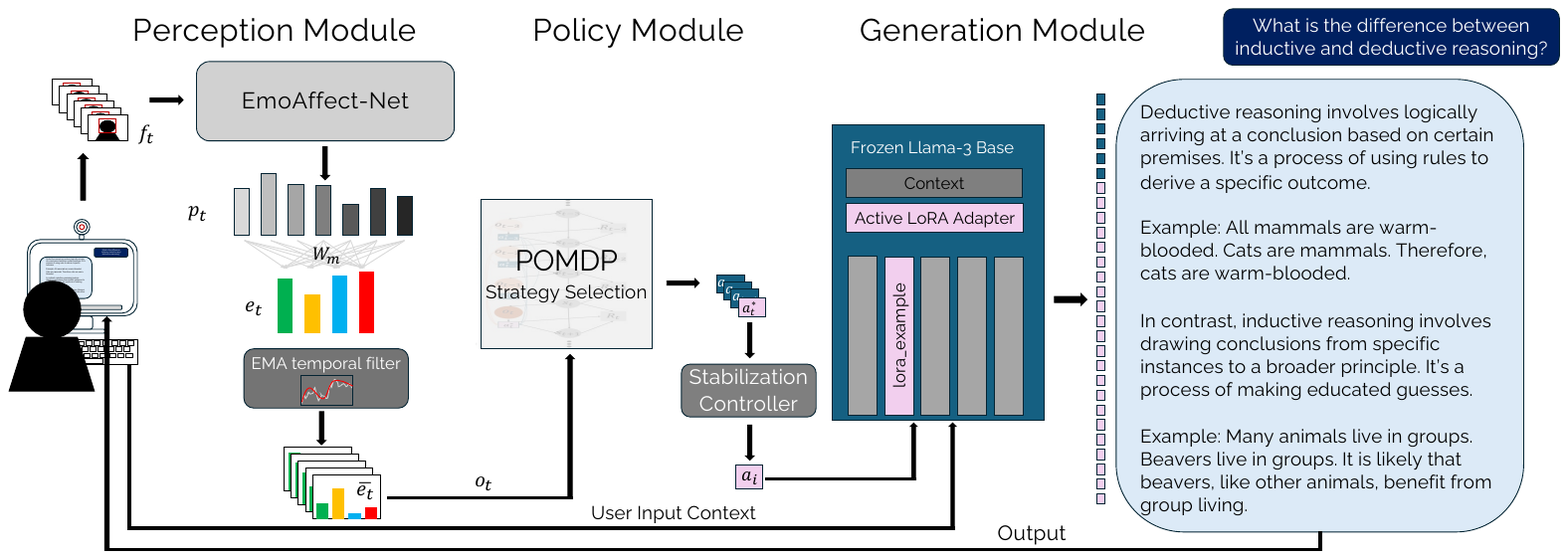}
  \caption{Overview of Aura's framework.
  During prompt--response interaction, our Perception Module maps the geometry of a person's facial expressions $f_t$ to robust observations $o_t$.
  The Policy Module performs belief tracking~$s_t$ and selects interventions~$a_i$ via a POMDP and Stabilization Controller.
  Aura's Generation Module then performs \emph{mid-turn shifts} during response generation via a suitable LoRA adapter following a frozen Llama-3 base.}
  \label{fig:overview}
\end{figure*}

\section{Related Work}

\subsection{Adaptive Language Generation and User State Modeling}

Recent work in NLP has explored adapting language generation to implicit user state, including latent background cues~\cite{jin2024implicit}, affective signals~\cite{pico2024exploring}, and unstated preferences~\cite{kim2025extracting}, in order to improve interaction quality and reduce conversational friction caused by system misunderstandings~\cite{mishra2023real, falke2020leveraging, ponnusamy2020feedback}.
Dialogue systems increasingly leverage signals such as dissatisfaction, confusion, or engagement to guide response selection, for example by predicting dissatisfaction and reranking outputs accordingly~\cite{see-manning-2021-understanding, jaques2019social, hancock2019learning}.
Prior work models attentive listener feedback and adaptive presentation strategies to dynamically adjust explanations during interaction~\cite{buschmeier2018communicative, axelsson-skantze-2019-modelling, yuan2024askingmultimodalclarifyingquestions}. 
However, such approaches typically operate at the turn level, selecting complete responses rather than modulating generation as it unfolds~\cite{wu2018response}.

Emotion-aware and empathetic response generation has become a prominent sub-area within ACL and EMNLP.
The EmpatheticDialogues benchmark shows that explicit emotion modeling improves perceived empathy and conversational quality, while subsequent work conditions generation on inferred emotion representations~\cite{rashkin-etal-2019-towards, lin2019moel, li2019empdg}.
MIME proposes mimicking the user’s emotional state to a controllable degree during generation, yielding improvements in both relevance and empathy~\cite{majumder-etal-2020-mime}.
Similarly, emotion conditioning is typically specified as a fixed control signal per response, following standard controllable generation or style conditioning approaches~\cite{keskar2019ctrl, zhang2024ctg}.
More recent systems integrate emotion across the dialogue pipeline, using inferred emotion to guide understanding and policy decisions such as elaboration or reassurance to improve task success and user experience~\cite{feng-etal-2024-infusing}.
Modeling latent user state under uncertainty has a long history in dialogue research through POMDPs, which maintain probabilistic belief states over user intent, goals, and observations~\cite{young2013pomdp}.
Extensions of this framework incorporate emotion and engagement-related states, enabling continuous belief updates over user emotion~\cite{dreamtochat2025}.
In contrast, our approach tightly couples continuous user emotional state inference with probabilistic state tracking and language generation, addressing settings where user state may shift during long or complex responses.
We maintain a POMDP belief over user state and dynamically activate multiple LoRA adapters within an LLM during decoding.
Building on recent advances in Parameter-Efficient Fine-Tuning PEFT and multi-adapter composition~\cite{wang2023lorahub}, the framework enables real-time modulation of generation without additional clarification turns.

\subsection{Emotion Recognition and Adaptive Interaction}
The broader idea of adapting system behavior based on user state has been studied in HCI~\cite{picard1997affective}.
Affective science distinguishes between \emph{emotion} as an observable context-dependent expression and broader \emph{affective state} as a latent psychological condition that unfolds over time~\cite{russell1980, barrett2019emotional}. 
Contemporary theories argue that facial expressions are imperfect and culturally mediated behavioral signals rather than direct readouts of internal mental states~\cite{barrett2019emotional}. 
Accordingly, we treat facial reactions as noisy behavioral proxies for interaction-relevant states. 
Prior work has shown that interfaces which respond to emotional cues can improve engagement, satisfaction, and task performance.
Emotion-aware user interfaces have leveraged sensors such as webcams to infer user emotions and adapt presentation or interaction style accordingly.
For example, emotion-aware messaging interfaces dynamically adjust UI elements based on detected facial expressions to better match the user's emotional state~\cite{kumaran2023echat}.
Empirical studies further demonstrate that facial signals captured during task execution correlate with post-task satisfaction and perceived system quality, and can be detected in real time using commodity hardware~\cite{pirmoradi2025bridging}.
Together, these findings establish the feasibility of unobtrusive, real-time sensing during natural human–computer interaction.

Emotion-aware dialogue systems use inferred user emotions to guide response selection or dialogue policy, often modeling emotional dynamics probabilistically~\cite{irfan2020dynamic}.
Surveys of emotionally adaptive agents show consistent gains in personalization and user satisfaction across domains~\cite{schlicher2025emotionally}.
However, most prior approaches adapt behavior externally to language generation, by selecting from predefined templates, adjusting stylistic labels, or choosing the next dialogue action at turn boundaries~\cite{yu-etal-2019-gunrock, keskar2019ctrl, zhou2017emotional, young2013pomdp}.
As a result, the inferred user state influences what response is chosen, but not how language is generated during a response.

\section{Method}
\topic{Our real-time framework modulates LLM inference based on inferred interaction states (Figure~\ref{fig:overview}) via three components:}
(1)~a Perception Module that infers facially expressed emotion distributions,
(2)~a Policy Module that selects interventions via a POMDP~\cite{kaelbling1998pomdp}, and
(3)~a Generation Module that executes interventions via LoRA adapters~\cite{hu2021lora}.
See Appendix~\ref{sec:appendix} for all details.

\subsection{Perception Module}
\topic{We employ a hierarchical pipeline to estimate interaction-centered user states from facially expressed emotions.}
We extract spatial features $\mathbf{f}_t$ from webcam frames following EmoAffect-Net via a ResNet-50 backbone pre-trained on AffectNet and process them through an LSTM over a temporal window ($N=10$) to produce a probability vector $\mathbf{p}_t$ across seven basic emotions: $\mathbf{p}_t = \text{softmax}(\mathbf{W}_h \text{LSTM}(\{\mathbf{f}_{t-k}\}) + \mathbf{b}_h)$
where $\mathbf{p}_t \in [0, 1]^7$ represents {Neutral, Happiness, Sadness, Surprise, Fear, Disgust, and Anger}~\cite{savchenko2021facial, mollahosseini2017affectnet, ekman1992argument}.

\subsubsection{State Mapping and Temporal Processing}
\topic{
Facial expression datasets rarely align with interaction-centered states. 
We map these probabilities into a four-dimensional state vector $\mathbf{e}_t \in \{\textbf{Engaged, Confused, Frustrated, Bored}\}$.}
Prior work identifies confusion, frustration, boredom, and engagement as primary predictors of disengagement and interaction breakdown~\cite{dmello2012dynamics, karumbaiah2022how}. 
We focus on interaction-relevant states directly actionable for adaptation. 
We therefore adopt a compact state space that balances interpretability, theoretical grounding, and computational tractability.
We refine a heuristic mapping empirically using the DAiSEE dataset to compute a Weighted Intensity Mapping matrix $\mathbf{W}_m$~\cite{gupta2016daisee}.
DAiSEE is well suited for this purpose as it provides large-scale, video-based labels for the four user states~\cite{gupta2016daisee}.
This yields the final state distribution $\mathbf{e}_t = \sigma(\mathbf{W}_m \mathbf{p}_t + \mathbf{b}_m)$ providing a distribution over intensities.
\topic{We apply exponential moving average (EMA) smoothing with spike detection to stabilize the POMDP observation $o_t$ against transient artifacts (e.g., blinks or occlusion) to prevent erratic policy interventions.}
The normalized vector $\bar{\mathbf{e}}_t$ serves as the formal observation.

\subsection{Policy Module}

\topic{We model intervention selection as a POMDP to manage uncertainty in inferred user states.} 
This framework enables principled belief tracking over noisy perceptual signals while maintaining a lightweight, real-time framework. 
The POMDP is defined by the tuple $(\mathcal{S}, \mathcal{A}, \mathcal{T}, \mathcal{R}, \Omega, \mathcal{O}, \gamma)$, where $\mathcal{S}$ represents user states (\textit{Engaged, Confused, Frustrated, Bored}) and $\mathcal{A}$ comprises communicative interventions (\textit{Reframe, Clarify, De-escalate, Pace, Simplify, Base}). 
User state evolution is governed by the transition function $\mathcal{T}(s' \mid s, a)$, which models the dynamics of state transitions in response to specific interventions. 
To account for the systematic noise inherent in the Perception Module, the observation model $\mathcal{O}(o \mid s)$ maps these latent states to perceptual evidence. 
The evidence lies in the observation space $\Omega$, a probability simplex where $o_t = \bar{\mathbf{e}}_t \in \Delta(\mathcal{S})$ gives the estimated probability distribution over the user's states at time $t$. 
Finally, the framework’s objectives are encoded through a reward function that assigns positive utility to Engaged states and penalizes Frustrated, Confused and Bored states, with a small action cost to discourage excessive intervention, and a discount factor $\gamma = 0.95$.
Incorrect or noisy observations do not immediately produce abrupt behavioral shifts. 
Because interventions require persistent evidence and hysteresis thresholds, false-positive detections might result only in temporary suboptimal strategy choices.

\subsubsection{Parameter Estimation}

\topic{We used GPT-4o to generate a synthetic corpus of 24,929 sessions (131,283~turns) to parameterize the POMDP model due to the scarcity of longitudinal data~\cite{openai2024gpt4o}.}
We enforced structured variance by: (i) pairing latent transitions ($s, s'$) with noisy observations ($o$) to enable accurate belief modeling; (ii) utilizing expert-guided prompting to identify optimal recovery strategies; and (iii) spanning 1–-12 turns across diverse topics and learner persistence levels for broad state-action coverage. 
The resulting dataset ($\mu=5.27 \pm 2.47$ turns/session) ensures policy robustness through significant representation of negative states, specifically Frustrated (18.28\%) and Bored (17.60\%).

\subsubsection{Optimization and Inference}
\topic{We estimate the coarse transition tendencies $\mathcal{T}$ and observation $\mathcal{O}$ models via Maximum Likelihood Estimation from the synthetic corpus using Laplace smoothing.}
\topic{At each timestep $t$, the framework maintains a belief state $b_t(s)$ by integrating the previous action $a_{t-1}$ and current observation $o_t$ via a recursive Bayesian update.} We perform this in log-space for numerical stability. To select the optimal intervention, we employ a QMDP approximation~\cite{littman1995learning}:
\begin{equation}
  \label{eq:lora}
  a_t^* = \arg\max_{a \in \mathcal{A}} \sum_{s \in \mathcal{S}} b_t(s) Q^*(s,a)
\end{equation}
where $Q^*(s,a)$ is a state--action value function computed once offline via tabular Q-learning ($\alpha = 0.1, \gamma = 0.95$)~\cite{watkins1992q}.
\topic{We implement a stabilization controller to ensure conversational continuity and prevent rapid ``flickering'' between strategies.}
This includes hysteresis (using enter/exit thresholds), a competitive counter mechanism, and a 3.0-second cooldown period.
This ensures that the interventions are intentional and persistent enough for the user to perceive.

\subsection{Generation Module}
\topic{Our Generation Module executes interventions selected by the Policy Module using modular, parameter-efficient adapters.}
Unlike full fine-tuning or heavy prompting, this dynamic activation allows mid-turn interventions while maintaining low-latency inference and preserving shared conversational context.

\begin{table}[ht]
\centering
\footnotesize
\begin{tabularx}{\columnwidth}{@{}X@{}}
\toprule
\textbf{Action (Adapter ID):} Strategy \\ \midrule
\textbf{Reframe} (\texttt{analogy}): Translates complex concepts into metaphors or analogies to bridge knowledge gaps. \\ \addlinespace[3pt]
\textbf{Clarify} (\texttt{example}): Provides concrete, real-world scenarios to ground abstract definitions. \\ \addlinespace[3pt]
\textbf{De-escalate} (\texttt{acknowledge}): Uses empathetic and supportive framing to reduce user frustration. \\ \addlinespace[3pt]
\textbf{Pace} (\texttt{summary}): Condenses previous content into core points to reduce information overload. \\ \addlinespace[3pt]
\textbf{Simplify} (\texttt{simplify}): Re-generates responses using basic vocabulary and shorter sentence structures. \\
\bottomrule
\caption{LoRA-based policy actions and strategies.}
\label{tab:action_mapping}
\end{tabularx}
\end{table}

\subsubsection{Intervention Library}

\topic{We implement five communicative interventions as LoRA grounded in research on communicative scaffolding, conversational grounding, and user-aware interaction~\cite{Sweller1988, Chi2001, Mayer2009, clark1991grounding, picard1997affective, rashkin-etal-2019-towards}.}
These strategies (Table~\ref{tab:action_mapping}) target common interaction challenges, including comprehension support (examples, analogies, simplification), pacing, and frustration mitigation.
We apply LoRA to a frozen Llama-3-8B-Instruct base~\cite{dubey2024llama}. For each linear projection $W_0 \in \mathbb{R}^{d \times k}$, the adapted forward pass is:
\begin{equation}
  \label{eq:lora}
  h = W_0 x + \Delta W x = W_0 x + \frac{\alpha}{r}(BA)x
\end{equation} with $r=16$, $\alpha=32$.
Adapters integrate into all linear layers ($W_q, W_k, W_v, W_o$, and MLP) across all blocks so that the model modulates discourse style while retaining shared linguistic knowledge.

\subsubsection{Strategy Distillation and Training}
\topic{We created a framework to generate synthetic data due to the lack of human-annotated datasets for fine-grained strategies.}
We first generate a neutral seed corpus $\mathcal{D}_{\text{seed}}$ of 500 educational question--answer interactions using Meta-Llama-3-8B-Instruct.
These interactions are neutral, clear, and factual, providing a baseline style suitable for transformation into distinct strategies. We transformed them into 2,500 strategy-specific variants using conditioned meta-prompts that emphasize distinct patterns.
We trained each adapter independently on its corpus via QLoRA (4-bit quantization) to optimize the autoregressive cross-entropy loss:\begin{equation}
  \label{eq:crossentropy}
  \mathcal{L}(\theta) = - \sum_{t=1}^{T} \log P(y_t \mid y_{<t}, x, \mathcal{C}, a_i; \theta)
\end{equation}where $\mathcal{C}$ is conversation history and $a_i$ is the active adapter~\cite{dettmers2023qlora}.
This ensures consistent strategy realization while remaining compatible with the shared base model.
Full meta-prompts and example outputs are provided in Table~\ref{tab:prompts} in the Supplementary Material.

\subsubsection{Dynamic Strategy Switching}
\topic{We developed a strategy-switching mechanism that enables mid-turn adaptation while preserving discourse continuity and shared conversational context.}
Strategy swaps occur only at sentence boundaries. When the policy triggers an intervention, the framework replaces active LoRA weights in less than 10\,ms.
Crucially, the framework preserves the Key--Value (KV) cache, allowing adapters to inherit conversational context without resetting discourse state or memory. 
Interventions alter presentation strategy while preserving semantic content through the shared frozen backbone.
To prevent brevity bias, a length-control mechanism ensures responses meet a minimum threshold ($L_{\min}=500$~chars).

\begin{table*}[t]
\centering
\small
\begin{tabular}{l cc cccc}
\toprule
& \multicolumn{2}{c}{\textbf{Effectiveness}} & \multicolumn{4}{c}{\textbf{Interaction Dynamics}} \\
\cmidrule(lr){2-3} \cmidrule(lr){4-7}
\textbf{Condition} & \textbf{Gain ($g$)} & \textbf{Acc. (\%)} & \textbf{Read. (\%)} & \textbf{Int. (s)} & \textbf{Task (s)} & \textbf{Turns} \\
\midrule
GPT-4o          & .54 ($\pm$.33)          & \textbf{90.4 ($\pm$6.0)} & 78.0 ($\pm$18.2) & 92.89 ($\pm$32.86)          & 47.53 ($\pm$24.01)          & 1.38 ($\pm$.49)          \\
Llama-3         & .41 ($\pm$.38)          & 89.8 ($\pm$7.8)          & 72.0 ($\pm$16.7)          & 99.88 ($\pm$41.91)          & 44.80 ($\pm$17.14)          & 1.35 ($\pm$.39)          \\
Aura (Ours)     & \textbf{.59$^*$ ($\pm$.22)} & 89.3 ($\pm$7.7)          & \textbf{78.0 ($\pm$16.4)} & \textbf{73.18$^{**}$ ($\pm$27.33)} & \textbf{44.74 ($\pm$19.86)} & \textbf{1.24 ($\pm$.23)} \\
\bottomrule
\end{tabular}
\caption{Learning outcomes and interaction dynamics. Results are reported as Mean ($\pm$SD). Bold denotes best performance; $^*p < .05$, $^{**}p < .01$ via Wilcoxon signed-rank test vs.\ Llama-3 baseline.}
\label{tab:performance_metrics}
\end{table*}

\begin{table*}[t]
\centering
\small
\setlength{\tabcolsep}{3pt}
\begin{tabular}{l cccc cccc}
\toprule
& \multicolumn{4}{c}{\textbf{Retrospective Evaluation}} & \multicolumn{4}{c}{\textbf{Per-Task Evaluation}} \\
\cmidrule(lr){2-5} \cmidrule(lr){6-9}
\textbf{Model} & \textbf{Understand.} & \textbf{Helpfulness} & \textbf{Accuracy} & \textbf{Predictability} & \textbf{Satisfaction} & \textbf{Clarity} & \textbf{Relevance} & \textbf{Confidence} \\
\midrule
GPT-4o          & 4.30 ($\pm$.57)          & \textbf{4.20 ($\pm$.70)} & 4.05 ($\pm$.60)          & 2.85 ($\pm$1.1)           & 3.70 ($\pm$1.1)           & \textbf{3.82 ($\pm$1.1)}  & 4.06 ($\pm$1.1)           & 3.72 ($\pm$1.0)           \\
Llama-3         & 4.20 ($\pm$.62)          & 4.05 ($\pm$.69)          & 3.95 ($\pm$.60)          & \textbf{2.95 ($\pm$1.2)} & 3.33 ($\pm$.92)          & 3.33 ($\pm$.88)          & 3.84 ($\pm$.91)          & 3.49 ($\pm$.82)          \\
Aura (Ours)     & \textbf{4.35 ($\pm$.49)} & 4.10 ($\pm$.55)          & \textbf{4.05 ($\pm$.60)} & 2.85 ($\pm$.99)          & \textbf{3.78 ($\pm$.87)} & 3.73 ($\pm$.88)          & \textbf{4.14 ($\pm$.89)} & \textbf{3.76 ($\pm$.82)} \\
\bottomrule
\end{tabular}
\caption{Subjective quality and performance metrics. Results are reported as Mean ($\pm$SD). Bold denotes best performance; $^*p < .05$ via Wilcoxon Signed-Rank test vs. Llama-3 baseline.}
\label{tab:quality_metrics}
\end{table*}

\section{Experimental Setup}
\label{sec:user_study}

We conducted a within-subject study ($N=20$) comparing Aura against non-adaptive Llama-3-8B and GPT-4o baselines. 
We selected this design to reduce inter-participant variability and increase sensitivity to interaction differences, while a Latin square design counterbalanced condition order and task blocks to control for learning and fatigue effects.
Aura is a closed-loop system in which each module jointly determines behavior; removing any component reduces it to the evaluated baseline. 
We evaluated adapters using Prometheus-Eval (Appendix C.2) to validate individual quality, confirming strong overall performance.

\subsection{Apparatus and Interfaces}
We conducted sessions in a standardized lab setting using a Logitech BRIO (30\,FPS) and an NVIDIA GeForce RTX 4090. 
To ensure equivalent end-to-end latency, all systems used the same web interface configured with a temperature of 0.7 and a per-character streaming delay of 0.05\,.

\subsection{Experimental Tasks and Protocol}
We designed 18 binary classification tasks to induce controlled cognitive friction, requiring participants to differentiate frequently conflated concepts across domains such as statistics, moral philosophy, and typography.
We normalized task difficulty using heuristics and independent LLM assessments, then distributed across three balanced blocks. 
To focus on conceptual understanding rather than example recall, participants were encouraged to reason through tasks conceptually rather than relying solely on memorized examples.
Tasks followed a three-phase cycle: (1) Pre-task: initial example and 7-point Likert self-report of understanding; (2) Interaction: participants engaged in a dialogue to get the information which concluded once the participant felt ready to answer similar questions; (3) Application: they classified four novel instances and reported their post-task understanding, allowing for the computation of the normalized learning gain $g = (\text{post} - \text{pre}) / (7 - \text{pre})$~\cite{Hake1998}.

\subsection{Evaluation Metrics}
We captured objective learning outcomes using binary task accuracy and normalized learning gain, alongside subjective measures such as post-task self-efficacy and confidence, to assess the effectiveness of adaptive interventions. 
Interaction dynamics were measured via total turn count, interaction duration, task completion duration, and the self-reported proportion of content consumed, providing insight into how adaptations influence conversation structure.  
User experience and output quality were evaluated through post-task questionnaires capturing satisfaction, perceived usefulness, and perceived quality of generated content.  
We measured perceived mental demand and interaction effort using items adapted from the NASA-TLX, while specific states (frustration, confusion, engagement, and boredom) were assessed via self-report scales~\cite{HartStaveland1988,Ceaparu2004, dmello2012dynamics}.
Valence and arousal were further captured using the Self-Assessment Manikin (SAM) and mapped to Russell's circumplex model of affect, allowing us to evaluate the impact of interventions on affective states~\cite{BradleyLang1994, russell1980}.

\subsection{Participants}
We recruited twenty participants (9~M, 11~F; mean age $M = 26.05$, $SD = 3.17$).
All participants reported high English proficiency ($M = 4.53/5.0$) and comfort with new technologies ($M = 4.40/5.0$).
Participants reported using chatbots/digital assistants daily or weekly and reported familiarity with LLMs ($M = 4.50/5.0$).

\begin{table*}[t]
\centering
\small
\begin{tabular}{l ccccc}
\toprule
& \multicolumn{5}{c}{\textbf{Affect \& Load: Post-Condition Retrospective}} \\
\textbf{Model} & \textbf{Valence} & \textbf{Arousal} $\downarrow$ & \textbf{Engagement}$^\dagger$ & \textbf{Demanding} $\downarrow$ & \textbf{Effort} $\downarrow$ \\
\midrule
GPT-4o          & \textbf{3.50 ($\pm$.89)} & 2.30 ($\pm$.86)          & 3.45 ($\pm$.83)          & 2.20 ($\pm$1.06)          & 2.30 ($\pm$1.22)          \\
Llama-3         & 3.35 ($\pm$.81)          & 2.35 ($\pm$.81)          & 2.85 ($\pm$.88)          & 2.35 ($\pm$.93)          & 2.55 ($\pm$1.05)          \\
Aura (Ours)     & 3.45 ($\pm$.69)          & \textbf{2.20 ($\pm$.62)} & \textbf{3.45$^*$ ($\pm$.60)} & \textbf{2.15 ($\pm$.93)} & \textbf{2.10 ($\pm$.97)} \\
\bottomrule
\end{tabular}
\vspace{0.75em}
\begin{tabular}{l cccccc}
\toprule
& \multicolumn{6}{c}{\textbf{Affect \& Experience: Per-Task Aggregated States}} \\
\textbf{Model} & \textbf{Frustration} $\downarrow$ & \textbf{Confusion} $\downarrow$ & \textbf{Boredom} $\downarrow$ & \textbf{Engagement} & \textbf{Valence} & \textbf{Arousal} $\downarrow$ \\
\midrule
GPT-4o          & 1.04 ($\pm$.62)          & 1.09 ($\pm$.67)          & 1.02 ($\pm$.61)          & 3.43 ($\pm$1.08)          & 3.20 ($\pm$.95)          & 1.04 ($\pm$.60)          \\
Llama-3         & 1.10 ($\pm$.52)          & 1.24 ($\pm$.46)          & .96 ($\pm$.43)          & 3.38 ($\pm$.85)          & 3.12 ($\pm$.91)          & 1.18 ($\pm$.53)          \\
Aura (Ours)     & \textbf{.95 ($\pm$.41)} & \textbf{1.04 ($\pm$.48)} & \textbf{.83 ($\pm$.41)} & \textbf{3.60 ($\pm$.90)} & \textbf{3.37$^*$ ($\pm$.86)} & \textbf{1.00 ($\pm$.53)} \\
\bottomrule
\end{tabular}
\caption{User states and perceived cognitive load. Results are reported as Mean ($\pm$SD). $\downarrow$ indicates lower is better. $^*p < .05$ via Wilcoxon Signed-Rank test vs. Llama-3 baseline; $^\dagger p < .05$ via Friedman test.}
\label{tab:affect_results}
\end{table*}
\section{Results}

We evaluated Aura across four dimensions: effectiveness, efficiency, quality, and affect.
We normalized all self-reported measures (e.g., frustration) by participants’ task understanding to control for topic difficulty.
We did not weight objective metrics.
We recorded per-task metrics immediately post-task, and conducted retrospective evaluations following each six-task block.

\subsection{Learning Effectiveness \& Interaction Dynamics}

As detailed in Table~\ref{tab:performance_metrics}, Aura yielded the highest normalized learning gain ($g=0.59\pm0.22$) compared to GPT-4o ($0.54\pm0.33$) and Llama-3 ($0.41\pm0.38$), and significantly outperformed the latter ($p = .028$, Wilcoxon; global Friedman $p = .167$).
Participants achieved consistently high accuracy across conditions (GPT-4o: 90.4\%; Llama-3: 89.8\%; Aura: 89.3\%).
The Aura condition significantly reduced interaction time ($73.18\text{s} \pm 27.33$) compared to GPT-4o ($92.89\text{s}$) and Llama-3 ($99.88\text{s}$), accounting for model latency, streaming behavior, turn length difference, and participant comprehension speed (Table~\ref{tab:performance_metrics}).
A Friedman test confirmed a significant global effect ($\chi^2=12.90, p=.002, W=0.323$), with post-hoc Wilcoxon tests showing a significant reduction against Llama-3 ($p < .001$).
While turn counts ($1.24$ vs.\ $1.35$ and $1.38$ for Aura, Llama-3, and GPT-4o), task time, and perceived reading proportion showed directional improvements, they did not reach statistical significance.

\subsection{User Experience and Output Quality}

Retrospective evaluations (Table~\ref{tab:quality_metrics}) indicated that Aura was perceived as the most effective for knowledge acquisition, leading in Understanding ($4.35$). Notably, while participants perceived identical factual Accuracy ($4.05$) between Aura and GPT-4o, Aura tied with GPT-4o on Predictability ($2.85$).
Participants were explicitly asked whether they noticed abrupt shifts, adaptive behavior, or response-style changes within any system. 
No participant reported perceiving abrupt transitions or mid-response adaptation in any of the systems.
Per-task ratings showed that Aura achieved the highest Satisfaction ($3.78$), Relevance ($4.14$) and Confidence ($3.76$) compared to both baselines.

\subsection{Affective State and Perceived Effort} 

As shown in Table~\ref{tab:affect_results},  Aura matched GPT-4o for the highest Engagement ($3.45$, $p < .05$ vs.\ Llama-3) while reporting the lowest Arousal ($2.20$), Demand ($2.15$), and Effort ($2.10$).
The Aura condition directionally reduced Boredom ($0.83$) versus Llama-3 ($0.96$) and maintained the highest task-level Engagement ($3.60$).
The reductions in Frustration ($0.95$) and Confusion ($1.04$) were directionally favorable, however not statistically significant.

\subsection{Participant Preference and Qualitative Feedback}

Post-study rankings showed a preference for Aura: 11 of 20 participants ranked it first, compared to 4 for GPT-4o and 5 for Llama-3, consistent with observed gains in efficiency, clarity, and engagement.
Participants described Aura as “concise and effective,” and producing the “least amount of overload,” mentioning a “no-nonsense” style, though one participant found its use of analogies “weird".
Individual feedback on GPT-4o emphasized its “clarity and detail,” with responses described as “easy to understand,” even for unfamiliar topics;
others reported “unpredictable” or “overwhelming” detail, ignored stylistic instructions (e.g., ELI5), overly complex language, or “too much information.”
Comments on Llama-3 varied: some participants noted “good structure” and “helpful summaries,” while others criticized inconsistent depth, ranging from “confusingly long” to “missing relevant information,” a perceived “lack of memory” that felt “random,” and heavy use of bullet points that made it “harder to maintain attention.”

\section{Discussion}

Our results demonstrate that Aura strikes a strategic balance between depth and brevity, yielding the highest normalized learning gains while reporting the lowest perceived mental demand and effort among the three conditions (Table~\ref{tab:affect_results}). 
Crucially, comparable perceived and quiz accuracy across chatbots, contrasted with directionally higher satisfaction for Aura and significantly higher retrospective engagement relative to Llama-3, highlights that the framework's advantage is not primarily a function of the underlying LLM's knowledge base.
Instead, these findings suggest that a friction point in current LLM interactions is a lack of communicative alignment.
By adjusting to the user's emotional state, the framework improves interaction efficiency without compromising learning outcomes, reducing mean interaction time by $\approx$21\% relative to GPT-4o and Llama-3. 
This "communicative tailoring" allows users to reach task proficiency faster than static, high-detail baselines by focusing on \textit{how} a model explains as much as \textit{what} it explains. 
Lower load-related retrospective ratings indicate that mid-turn modulation allows users to complete tasks with reduced mental exertion, demonstrating improved efficiency. 
The "information overload" reported in the GPT-4o condition, specifically its tendency to ignore style instructions (e.g., ELI5), and Llama-3's "inconsistent" depth forced participants to work harder to extract relevance.
In contrast, Aura's delivery maintained high engagement by providing depth better aligned to the immediate context. 
Directional reductions in boredom and frustration, alongside increased engagement, suggest that modulating delivery may mitigate interaction fatigue. 
By preemptively addressing cognitive disequilibrium, the POMDP automates the explicit repair process, contributing to shorter interactions and more sustained engagement. 
Rather than forcing users to realize confusion and formulate clarifying prompts, the framework intervenes before the conversational flow breaks. 
Comparable predictability and perceived accuracy across conditions, alongside participants reporting no noticeable abruptness or adaptation, suggest that intra-turn modulation did not manifest as unreliable behavior. We attribute this stability to probabilistic belief filtering and stabilization mechanisms that suppress transient noise and prevent erratic switching, while preserved conversational context enables smooth and coherent adapter transitions.

\section{Conclusion}
We presented Aura, a real-time adaptive LLM framework that modulates response strategies based on inferred user state derived from facially expressed signals. 
Combining POMDP-based policy selection with dynamic LoRA activation, Aura moves beyond static prompt--response exchanges toward communicative alignment.
In a user study, this approach reduced interaction time compared to GPT-4o and Llama-3 while achieving highest learning gains. 
Participants overall preferred our framework, citing its conciseness and reduced overload.  
These results demonstrate that conversational AI effectiveness depends not only on information quality but on the model's ability to synchronize its delivery with the user’s immediate context.

\section{Limitations}

While Aura demonstrates potential for intra-turn adaptation, several factors bound the generalizability of our findings. 
First, our reliance on webcam-based facial cues may miss subtle user states or those expressed through other modalities, such as vocal prosody, gaze trajectory, or typing rhythm. 
Furthermore, facial emotion models and mappings are subject to cultural bias, as emotional expression and instructional preferences vary globally.
Moreover, facially expressed emotions should be interpreted cautiously: facial emotion recognition remains an actively debated area, and such signals are imperfect behavioral proxies that may not reliably reflect internal affective or cognitive states~\cite{barrett2019emotional}.

From a technical standpoint, the POMDP transition and reward models were learned from a synthetic corpus; while this enables controlled policy learning, it may limit coverage of the more complex dynamics found in the wild. Furthermore, adaptation is currently restricted to a discrete set of LoRA-based rhetorical strategies, which simplifies the more continuous and nuanced stylistic shifts characteristic of human communication.

Our evaluation was also limited to binary classification tasks designed to induce controlled cognitive friction. 
Future work should investigate intra-turn adaptation in open-ended or multi-step tasks, where success depends on subjective preferences rather than objective correctness.
Finally, as noted in our qualitative feedback, proactive interventions based on implicit cues can feel "weird" to users by bypassing explicit agency.

\section{Ethical Considerations}

The primary ethical concern in this work involves the collection and processing of facial geometry data. 
To mitigate privacy risks, we ensured all facial landmark extraction was performed locally and in real time, meaning no raw video footage or identifiable biometric images were stored or transmitted to external servers. 
The framework operates only on transient, high-level emotion probability estimates used for short-term interaction adaptation.

However, we acknowledge that facial emotion recognition is prone to algorithmic bias, where facial expression models may fail to generalize across different ethnicities, genders, ages, or cultural backgrounds. 
Because expressions of confusion or frustration are not universal, our framework risks misinterpreting users whose expressive norms differ from the training data, including individuals with diminished facial expressivity or those with specific disabilities like facial paralysis. 

Finally, all participants (mostly students recruited from the university) in our study provided written informed consent, were fully briefed on the sensing technology being utilized.
Participants were compensated with a small, non-monetary token and participated under a protocol ensuring transparency, privacy, and voluntary consent.

\balance
\bibliography{custom}

\clearpage
\appendix

\section{Perception Module \& Signal Processing}
\label{sec:appendix}
\subsection{The DAiSEE Dataset: Rationale for Selection} We utilize the Dataset for Affective Analysis in E-learning Environments (DAiSEE)~\cite{gupta2016daisee} specifically because it bridges the gap between general computer vision and educational psychology. Unlike traditional datasets such as CK+ or FER2013, which focus on universal basic emotions (e.g., "Disgust" or "Fear"), DAiSEE is grounded in pedagogical contexts. It contains over 9,000 video snippets annotated with four interaction-centered states: Boredom, Confusion, Engagement, and Frustration. Labels are provided on a four-point ordinal scale ranging from 0 (neutral) to 3 (high).

\subsection{Empirical Refinement of State Mapping}
To bridge the gap between basic facial expressions and interaction states, we derive the mapping matrix $\mathbf{W}_m$ empirically. Each weight $w_{jk}$ represents the expected intensity of an interaction state $j$ given the probability of a basic emotion $k$. Using $D$ samples from DAiSEE, the weights are calculated as:

\begin{equation}
w_{jk} = \frac{\sum_{i=1}^{D} L_{i,j} \cdot p_{i,k}}{\sum_{i=1}^{D} p_{i,k}}
\end{equation}

where $L_{i,j}$ is the ground-truth intensity for state $j$ in sample $i$, and $p_{i,k}$ is the inferred probability of basic emotion $k$. This ensures the transition from facial geometry to pedagogical states is data-driven and robust to sensor variance.

\subsection{Robustness via Nonlinear Smoothing}
To mitigate sensor artifacts (e.g., blinks) while maintaining state persistence, we employ a dual-mode filter. Given a spike-detection threshold $\beta = 1.4$ and a smoothing factor $\gamma = 0.8$, the filtered state vector $\bar{\mathbf{s}}_t$ is updated as:

\begin{equation}
\bar{\mathbf{s}}_t = 
\begin{cases} 
\frac{\bar{\mathbf{s}}_{t-1} + \mathbf{s}_t}{2} & \text{if } \mathbf{s}_t > \bar{\mathbf{s}}_{t-1} \cdot \beta \\
\gamma \bar{\mathbf{s}}_{t-1} + (1 - \gamma) \mathbf{s}_t & \text{otherwise}
\end{cases}
\end{equation}

\section{Policy Module Details}

\subsection{Transition Model Estimation}
We derive the transition model $\hat{\mathcal{T}}(s' \mid s, a)$ from the synthetic corpus using transition frequencies. To handle data sparsity, we apply Laplace smoothing ($\kappa = 1$):
\begin{equation}
\hat{\mathcal{T}}(s' \mid s, a) = \frac{N(s,a,s') + \kappa}{\sum_{s'' \in \mathcal{S}} (N(s,a,s'') + \kappa)}
\end{equation}
where $N(s,a,s')$ is the count of transitions from state $s$ to $s'$ under action $a$.

\subsection{Recursive Belief Tracking}

The framework updates the belief state $b_t(s')$ by integrating the current observation $o_t$ (from the Perception Module) with the transition dynamics and the previous action $a_{t-1}$:
\begin{equation}
b_t(s') = \eta \, \mathcal{O}(o_t \mid s') \sum_{s \in \mathcal{S}} \mathcal{T}(s' \mid s, a_{t-1}) b_{t-1}(s)
\end{equation}
where $\mathcal{O}(o_t \mid s')$ is the observation model and $\eta$ is a normalization constant.

\subsection{Offline Policy Optimization}
The state-action value function $Q(s,a)$ is optimized using tabular Q-learning on the synthetic environment following the temporal difference (TD) update logic:
\begin{equation}
Q(s,a) \leftarrow (1-\alpha)Q(s,a) + \alpha [r + \gamma \max_{a'} Q(s',a')]
\end{equation}

where $\alpha = 0.1$ and $\gamma = 0.95$. This yields the $Q^*(s,a)$ values used for real-time QMDP action selection.

\subsection{Synthetic Trajectories with Belief States}
Table~\ref{tab:belief_traces} illustrates a few examples of the synthesized trajecories including True State, the framework's internal Belief Distribution $b_t$, and the resulting Action. The belief vector represents probabilities for [B]ored, [C]onfused, [F]rustrated, and [E]ngaged.

\begin{table*}[ht]
\small
\centering

\begin{tabularx}{\textwidth}{@{}l c c l X r @{}}
\toprule
\textbf{Topic} & \textbf{True State} & \textbf{Belief Vector} $[B, C, F, E]$ & \textbf{Action} & \textbf{Dialogue Trace} & \textbf{Rew.} \\ \midrule
\textit{Neural Networks} & Confused & $[0.08, \mathbf{0.57}, 0.12, 0.23]$ & Clarify & \textbf{U:} "I don't get backprop." \newline \textbf{C:} "Think of it as step-by-step error reduction." & +14 \\
 & Engaged & $[0.03, 0.09, 0.04, \mathbf{0.84}]$ & Base & \textbf{U:} "Oh, now I see how errors guide learning!" \newline \textbf{C:} "Exactly! Each step fine-tunes the weights." & +11 \\ \midrule
\textit{Climate Change} & Bored & $[\mathbf{0.52}, 0.18, 0.13, 0.17]$ & Reframe & \textbf{U:} "This is getting a bit dull." \newline \textbf{C:} "Think how sea levels affect your specific city." & +6 \\
 & Frustrated & $[0.11, 0.14, \mathbf{0.67}, 0.08]$ & De-escalate & \textbf{U:} "It's too much information to grasp!" \newline \textbf{C:} "Let's focus only on Arctic ice melting for now." & +13 \\ \midrule
\textit{History of AI} & Bored & $[\mathbf{0.58}, 0.12, 0.07, 0.23]$ & Reframe & \textbf{U:} "The old AI history seems kind of dry." \newline \textbf{C:} "Consider how it inspired the self-driving cars of today!" & +9 \\
\bottomrule

\end{tabularx}
\caption{Example synthetic trajectories}
\label{tab:belief_traces}
\end{table*}

\section{Generation Module \& LoRA Details}

\subsection{Strategy Distillation Meta-Prompts}
To create the strategy-specific training data, we utilized five distinct meta-prompts to transform neutral educational content into interventions (Table~\ref{tab:prompts}). The goal was to preserve the factual accuracy of the base model while the style is modulated.

\begin{table*}[ht]
\small
\centering

\begin{tabularx}{\textwidth}{@{}l X @{}}
\toprule
\textbf{Adapter} & \textbf{Conditioning Meta-Prompt} \\ \midrule
\textit{Reframe} & "Reframe the following answer using a clear, engaging analogy. Keep all important steps, but explain it in a new, metaphorical way." \\ \midrule
\textit{De-escalate} & "Rewrite the answer to be more encouraging and supportive. Validate the user's difficulties indirectly, but NEVER mention feelings explicitly (e.g., do not say 'you might feel frustrated')." \\ \midrule
\textit{Clarify} & "Rewrite the answer by providing a clear, tangible example that illustrates the abstract points or steps to make it concrete." \\ \midrule
\textit{Simplify} & "Rewrite the following answer using simpler vocabulary and shorter, clearer sentences while keeping all important information." \\ \midrule
\textit{Pace} & "Summarize the following answer concisely, keeping the most important information. Use bullets or numbered lists if needed." \\
\bottomrule
\end{tabularx}
\caption{Meta-Prompts for LoRA Strategy Distillation}
\label{tab:prompts}
\end{table*}

\subsection{Adapter Evaluation Details}
\begin{table}[h]
\small
\centering
\begin{tabularx}{\columnwidth}{l X}
\toprule
\textbf{Metric} & \textbf{Survey Item / Question} \\ 
\midrule
\textit{Satisfaction} & I feel satisfied with my interaction with the chatbot. \\
\textit{Frustration} & I felt frustrated while performing this task. \\
\textit{Confusion} & I felt confused while performing this task. \\
\textit{Boredom} & I felt bored while performing this task. \\
\textit{Engagement} & I felt engaged while performing this task. \\
\textit{Valence} & Overall I felt... (Very Negative to Very Positive). \\
\textit{Arousal} & I felt... (Very Calm to Very Tense). \\
\textit{Clarity} & The output was clear and easy to understand. \\
\textit{Relevance} & The output was relevant to my question/task. \\
\textit{Cognitive Load} & The mental effort required to understand the output was high. \\
\textit{Confidence} & I feel confident in my ability to complete the task accurately. \\
\bottomrule
\end{tabularx}
\caption{Subjective evaluation metrics and corresponding survey items used in the post-task assessment.}
\label{tab:survey_metrics}
\end{table}

Table~\ref{tab:prometheus_eval} presents the zero-shot evaluation of our LoRA adapters using the Prometheus-eval framework~\cite{kim2024prometheus}. Each adapter was evaluated on a 5-point Likert scale across nine functional dimensions to ensure that specialization did not degrade core capabilities.
\begin{table*}[ht]
\centering
\small
\begin{tabular}{l cccccccccc}
\toprule
\textbf{Adapter} & \textbf{GR} & \textbf{IF} & \textbf{PL} & \textbf{RE} & \textbf{RF} & \textbf{SA} & \textbf{ToM} & \textbf{TU} & \textbf{ML} & \textbf{Overall} \\
\midrule
Base (Llama-3)   & 4.15 & 4.11 & 4.34 & 3.58 & 3.67 & 4.16 & 4.07 & 3.40 & 3.43 & 3.88 \\
\texttt{lora\_acknowledge}  & 4.18 & \textbf{4.36} & \textbf{4.40} & 3.34 & \textbf{4.14} & 3.96 & 4.11 & 3.63 & 3.29 & \textbf{3.93} \\
\texttt{lora\_summary}  & 4.19 & 4.18 & 4.33 & \textbf{3.60} & 3.88 & \textbf{4.29} & 4.10 & \textbf{3.67} & 3.21 & \textbf{3.94}\\
\texttt{lora\_example} & \textbf{4.20} & 4.19 & 4.16 & 3.55 & 3.83 & 4.25 & 4.14 & 3.51 & \textbf{3.74} & \textbf{3.95}* \\
\texttt{lora\_analogy}  & 4.05 & 4.05 & 4.37 & 3.54 & 3.98 & 3.96 & \textbf{4.15} & 3.57 & 3.27 & \textbf{3.88} \\
\texttt{lora\_simplify} & 3.79 & 3.80 & 4.00 & 3.37 & 3.48 & 3.53 & 3.64 & 3.27 & 2.99 & 3.54 \\
\bottomrule
\end{tabular}
\caption{Adapter capabilities evaluated via \texttt{prometheus-eval} (1--5 scale). 
\textbf{GR}: Grounding, \textbf{IF}: Instruction Following, \textbf{PL}: Planning, \textbf{RE}: Reasoning, \textbf{RF}: Refinement, \textbf{SA}: Safety, \textbf{ToM}: Theory of Mind, \textbf{TU}: Tool Usage, \textbf{ML}: Multilingual.}
\label{tab:prometheus_eval}
\end{table*}

\section{Experimental Setup Details}

\subsection{Binary Classification Task Bank}
We designed 18 tasks across varying difficulty levels (1: Low to 3: High). Each task consists of a representative sample provided and four novel instances for user classification (Table~\ref{tab:task_bank_part1}, Table~\ref{tab:task_bank_part2}).

\subsection{User Study Interface}

\begin{figure*}[t]
  \includegraphics[width=\textwidth]{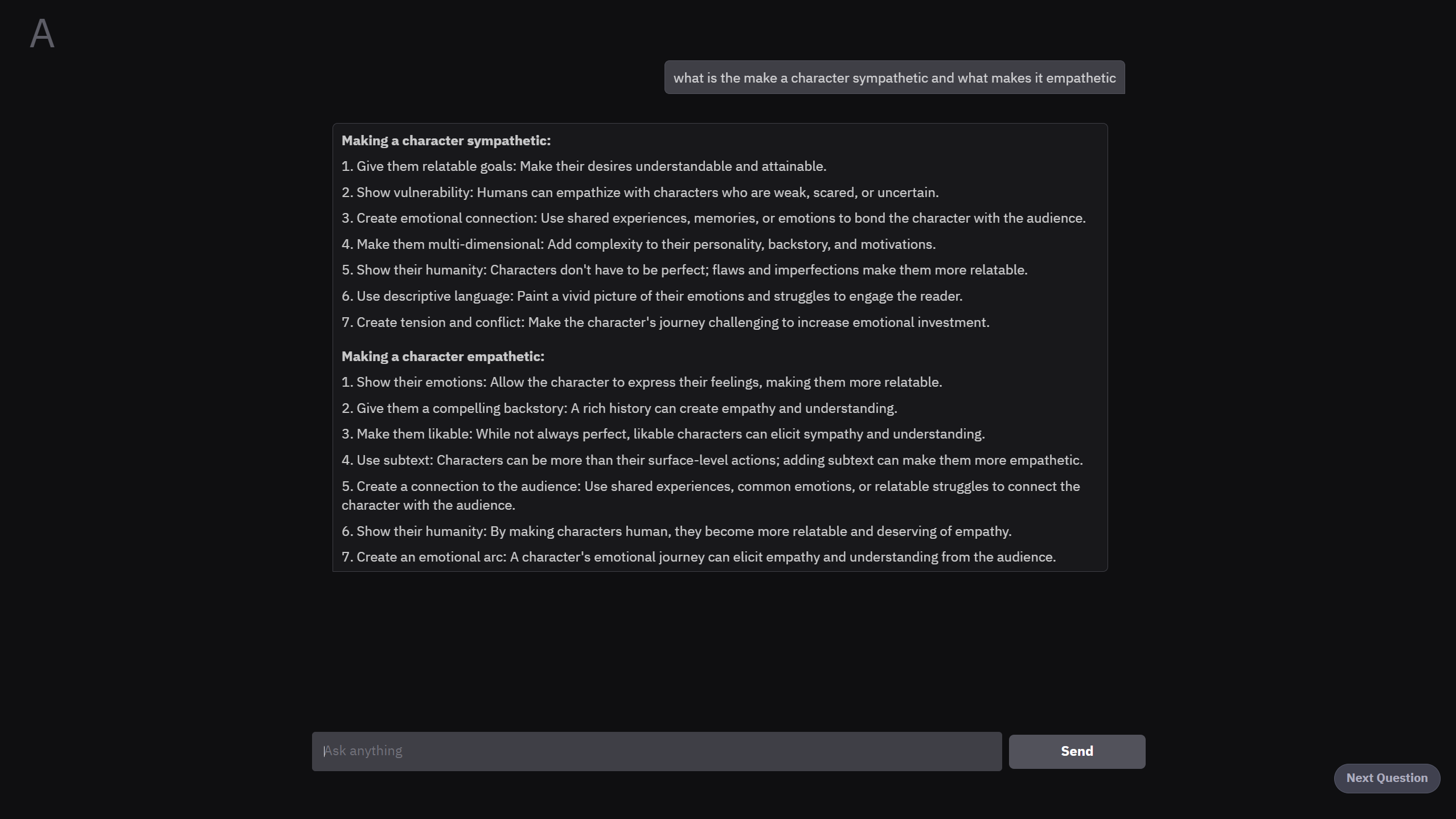}
  \caption{Experimental apparatus and interface layout. The chat layout used during the user study for all three conditions}
  \label{fig:chatinterface}
\end{figure*}

Participants interacted with the web-based chat interface illustrated in Figure~\ref{fig:chatinterface}. To ensure experimental validity and minimize visual bias, the UI layout remained identical across all three experimental conditions (Llama-3-8B Base, GPT-4o, and Aura). The chatbox served as the central hub for streaming LLM responses and user input. To ensure consistent interaction velocity, all responses were delivered using a synchronized character-by-character streaming mechanism. The participants were asked to read the output as it was generating.

\subsection{Assessment Interface}
\begin{figure}[t]
  \includegraphics[width=\columnwidth]{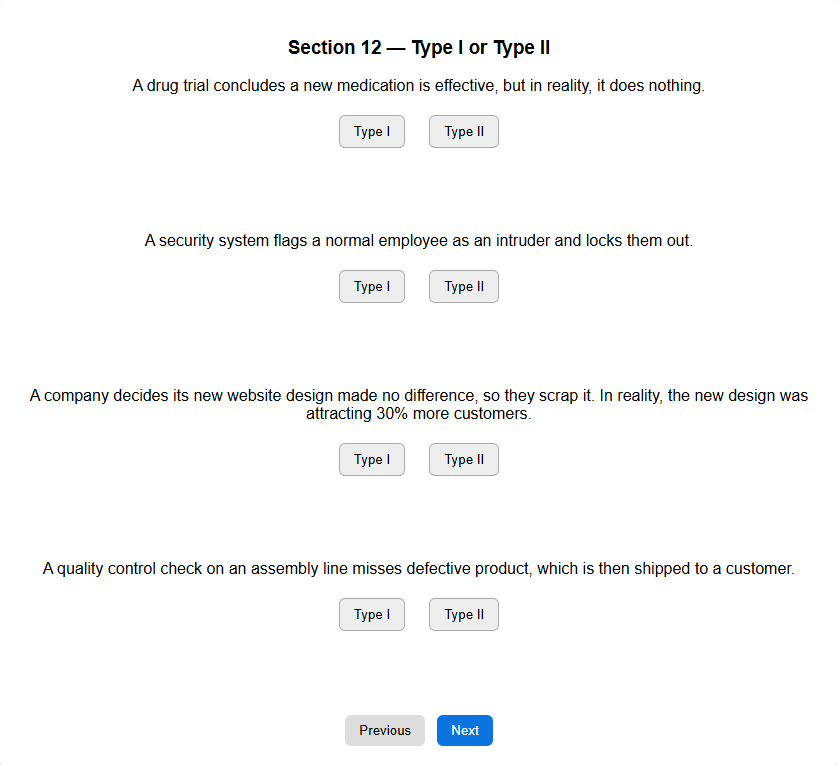}
  \caption{The binary classification task interface. Each screen presents a domain-specific problem statement designed to induce cognitive friction, followed by two competing hypotheses.}
  \label{fig:task}
\end{figure}

The binary classification interface, shown in Figure~\ref{fig:task}, shows an example of what the participants were presented with after completing their interaction with the chatbot.

\begin{table*}[ht]
\small
\centering
\caption{Classification Tasks (Tasks 1--9)}
\label{tab:task_bank_part1}
\begin{tabularx}{\textwidth}{@{}l c >{\hsize=0.8\hsize\RaggedRight\arraybackslash}X >{\hsize=1.3\hsize\RaggedRight\arraybackslash}X @{}}
\toprule
\textbf{Topic} & \textbf{Tier} & \textbf{Sample Question} & \textbf{Test Items} \\ \midrule
\textbf{Typography} & 1 & Coca-Cola & \textbullet~Google \newline \textbullet~The New York Times \newline \textbullet~Tiffany \& Co \newline \textbullet~Microsoft \\ \midrule
\textbf{Energy} & 1 & Wind Power & \textbullet~Coal \newline \textbullet~Natural Gas \newline \textbullet~Solar Power \newline \textbullet~Nuclear energy \\ \midrule
\textbf{Geology} & 1 & Shale forms from compressed mud layers in calm water. & \textbullet~Granite forms from cooled magma deep underground. \newline \textbullet~Limestone forms from compacted marine shells. \newline \textbullet~Sandstone develops from compressed sand layers. \newline \textbullet~Basalt forms from lava that cooled on the Earth's surface. \\ \midrule
\textbf{Biology} & 1 & A body cell divides to repair damaged tissue. & \textbullet~Produces gametes with half the number of chromosomes. \newline \textbullet~Occurs in reproductive organs to form sperm and eggs. \newline \textbullet~Results in two identical daughter cells. \newline \textbullet~Responsible for normal cell growth and repair. \\ \midrule
\textbf{Music} & 2 & A lullaby meant to calm a crying baby. & \textbullet~A triumphant anthem for a winning sport team. \newline \textbullet~A somber soundtrack for a detective investigating a crime scene. \newline \textbullet~A cheerful, upbeat song for a children's television show. \newline \textbullet~A dramatic, intense piece of music for a movie villain's entrance. \\ \midrule
\textbf{Literature} & 2 & A doctor who comforts a patient by understanding their fear. & \textbullet~A lonely orphan who is treated poorly but remains kind. \newline \textbullet~A wise old mentor who always knows how a friend is feeling. \newline \textbullet~A starship captain who excels at understanding alien cultures. \newline \textbullet~A down-on-their-luck detective trying to solve one last case. \\ \midrule
\textbf{Art History} & 2 & A landscape painting capturing the exact lighting of a sunset. & \textbullet~An artist aims to capture the exact quality of sunlight on a haystack at different times of day. \newline \textbullet~An artist paints a portrait with unnatural colors to show inner torment. \newline \textbullet~A painter uses soft, blended brushstrokes to show fog over a river. \newline \textbullet~A painter uses jagged lines to convey anxiety and despair. \\ \midrule
\textbf{Egyptian Hist.} & 2 & Akhenaten promotes worship of the sun disk Aten above all other gods. & \textbullet~The Great Pyramid of Giza is constructed. \newline \textbullet~Tutankhamun rules and restores traditional gods. \newline \textbullet~Hatshepsut commissions temple expansions and trade expeditions. \newline \textbullet~Ramses II leads military campaigns into Syria. \\ \midrule
\textbf{WWII History} & 2 & Germany launches a blitzkrieg campaign through Poland in 1939. & \textbullet~The Normandy invasion (D-Day) is planned and executed. \newline \textbullet~Operation Barbarossa launches a massive invasion of the USSR. \newline \textbullet~The Battle of Britain is fought primarily in the air over the UK. \newline \textbullet~The island-hopping campaign advances through the Pacific. \\
\bottomrule
\end{tabularx}
\end{table*}

\begin{table*}[ht]
\small
\centering
\caption{Classification Tasks (Tasks 10--18)}
\label{tab:task_bank_part2}
\begin{tabularx}{\textwidth}{@{}l c >{\hsize=0.8\hsize\RaggedRight\arraybackslash}X >{\hsize=1.3\hsize\RaggedRight\arraybackslash}X @{}}
\toprule
\textbf{Topic} & \textbf{Tier} & \textbf{Sample Question} & \textbf{Test Items} \\ \midrule
\textbf{Psychology} & 2 & A child salivates when hearing the bell that signals lunch. & \textbullet~A dog learns to sit because it receives a treat afterward. \newline \textbullet~A person feels anxious when entering a dentist's office. \newline \textbullet~A child fears a white rabbit after being frightened by one. \newline \textbullet~A student studies more because good grades bring praise. \\ \midrule
\textbf{Astronomy} & 2 & A distant galaxy’s light wavelength is stretched. & \textbullet~A star's spectral lines move toward shorter wavelengths. \newline \textbullet~A planet approaching Earth causes light to compress. \newline \textbullet~A galaxy receding from Earth shows longer wavelengths. \newline \textbullet~Supernova light appears stretched through expanding space. \\ \midrule
\textbf{Statistics} & 2 & A pregnancy test fails to detect a pregnancy that exists. & \textbullet~A drug trial concludes a medication is effective when it is not. \newline \textbullet~A security system flags a normal employee as an intruder. \newline \textbullet~A company scraps a website design that was 30\% better. \newline \textbullet~Quality control misses a defective product on an assembly line. \\ \midrule
\textbf{Archaeology} & 2 & A farmer uses a stone sickle to harvest early domesticated wheat. & \textbullet~A village shows evidence of simple stone structures. \newline \textbullet~A burial site contains bronze weapons and ornaments. \newline \textbullet~Large fortified settlements with metal tools appear. \newline \textbullet~Pottery is hand-shaped and undecorated. \\ \midrule
\textbf{Logic} & 2 & All humans are mortal. Socrates is human. Therefore, Socrates is mortal. & \textbullet~Every crow observed so far is black; therefore, all crows are black. \newline \textbullet~If it rains, the ground gets wet. It is raining. Therefore, it is wet. \newline \textbullet~This metal expands when heated; therefore, all metals expand. \newline \textbullet~Triangles have three sides. Therefore, this shape is a triangle. \\ \midrule
\textbf{Linguistics} & 2 & The sound /p/ distinguishes 'pat' from 'bat'. & \textbullet~The plural suffix ``-s'' in ``cats''. \newline \textbullet~Vowel change distinguishing ``bit'' and ``beat''. \newline \textbullet~The consonant /k/ in ``cat'' and ``kit''. \newline \textbullet~The prefix ``un-'' in ``unhappy''. \\ \midrule
\textbf{Economics} & 3 & The government reduces income taxes to increase consumer spending. & \textbullet~A central bank announces it is raising interest rates. \newline \textbullet~The government passes a law to increase infrastructure spending. \newline \textbullet~A new tax credit for families reduces taxes owed. \newline \textbullet~Central bank buys bonds to increase money supply. \\ \midrule
\textbf{Law} & 3 & A court strictly applies a written penal code to reach a verdict. & \textbullet~A lawyer argues a case by citing the rulings of previous cases. \newline \textbullet~A judge consults an article in a pre-written book of codes. \newline \textbullet~A supreme court decision becomes the standard for lower courts. \newline \textbullet~A legislature writes an exhaustive code from scratch. \\ \midrule
\textbf{Philosophy} & 3 & Lying to save a life is still wrong because lying breaks a moral rule. & \textbullet~Act is justified if it produces the greatest happiness. \newline \textbullet~Killing one to save five maximizes total good. \newline \textbullet~Stealing is wrong regardless of potential benefits. \newline \textbullet~Breaking a promise is acceptable if it prevents harm. \\
\bottomrule
\end{tabularx}
\end{table*}

\end{document}